\documentclass[aps,showpacs,twocolumn, tightenlines]{revtex4}
\usepackage{epsfig,amssymb,bm,graphics}

\begin{document}


\title{Comment on the $\Theta^+$-production at high energy}


 \author{A.I.~Titov$^{ab}$,  A. Hosaka$^{c}$, S. Dat\'e$^d$, and Y. Ohashi$^d$}
 \affiliation{
 $^a$Advanced Photon Research Center, Japan Atomic Energy Research Institute,
 Kizu, Kyoto
 619-0215, Japan\\
 $^b$Bogoliubov Laboratory of Theoretical Physics, JINR,
  Dubna 141980, Russia\\
 $^c$Research Center of Nuclear Physics, Osaka University,
  Ibaraki, Osaka 567-0047, Japan\\
 $^d$Japan Synchrotron Radiation Research Institute, SPring-8,
1-1-1 Kouto Mikazuki-cho Sayo-gun Hyogo 679-5198, Japan
 }


\begin{abstract}
We show that the cross sections of the $\Theta^+$-pentaquark production
in different processes decrease with energy faster than the cross
sections of production of the conventional three-quark hyperons.
Therefore, the threshold region with the initial energy of a few GeV
or less seems
to be more favorable for the production and experimental study of
$\Theta^+$-pentaquark.
\end{abstract}

 \pacs{13.60Le, 13.75.Jz,13.85.Fb}

 \maketitle

The discovery of the $\Theta^+$-pentaquark by the LEPS at
SPring-8~\cite{Nakano03} and its subsequent confirmation  in a
series of other
experiments~\cite{DIANA,CLAS1,CLAS2,SAPHIR,Asratyan} performed
mainly at low energy poses a problem of the energy dependence of
the $\Theta^+$-production cross section and inspires to make
further confirmation of other penta-quarks  at high energy. In
this Rapid Communication we analyze the high-energy limit of
$\Theta^+$-production in three kinematical regions: exclusive
production at large momentum transfers, exclusive production at
diffractive region and the $\Theta^+$-production in inclusive
processes in the fragmentation region. We show that at all cases
the  $\Theta^+$ production cross section is suppressed compared to
the production of the "conventional" three-quark hyperons. In
order to remove dimensional parameters we will consider ratios of
cross sections taken at two energies: relatively small energy
$s_0$ (which is the reference point) and at large energy $s$
\begin{eqnarray}
  R_Y=\frac{d\sigma_Y(s)}{d\sigma_Y(s_0)}, \qquad
  R_\Theta=\frac{d\sigma_\Theta(s)}{d\sigma_\Theta(s_0)},
  \qquad R_{\Theta Y}=\frac{R_\Theta}{R_Y}.
\end{eqnarray}

\section{exclusive $\Theta^+$-production at large momentum transfers}

 For definiteness sake, let us consider the $\pi N\to \Theta^+\bar K$
 and $\pi N\to YK$ processes. The energy dependence of the
 invariant amplitude of
 exclusive hadronic process $AB\to CD$ at large momentum transfers
 has the "automodel" (scale)  behaviour~\cite{MMT,BF}
 which is defined by the dimension of the connected Born
 amplitudes. The corresponding diagrams for $\pi N\to YK$
 and $\pi N\to \Theta^+\bar K$ reactions are shown in Figs.~1(a) and (b),
 respectively.
 \begin{figure}[ht] \centering
 \includegraphics[width=0.55\columnwidth]{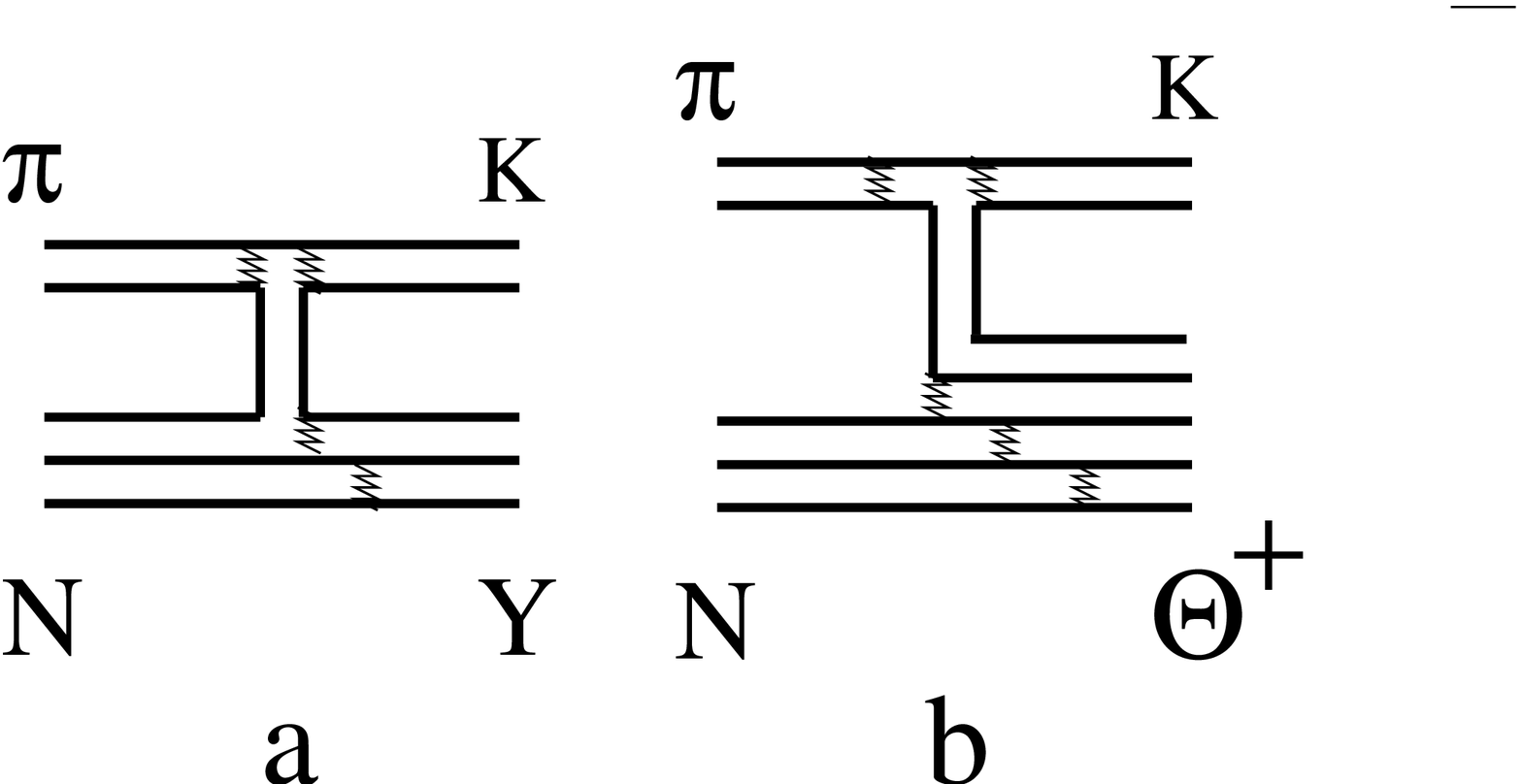}
 \caption{The Born diagrams for the $\pi N\to YK$ (a)
 and $\pi N\to \Theta^+\bar K$ (b) reactions at large momentum transfers.}
 \label{fig:1}
 \end{figure}
The dimension of the invariant amplitude $T_{\pi N\to YK}$ is
 \begin{eqnarray}
 [\rm length]^{n_{\pi}+n_N+n_K+n_Y-4}
 =[\rm length]^{6}.
 \end{eqnarray}
 where ${n_h}$ is a minimal number of the constituent (quarks)
 in a hadron $h$ involved in the process and
 the Dirac spinor are normalized
 as $\bar u(p)\gamma_\alpha u(p)=2p_\alpha$. This results in power
  decreasing of the corresponding cross sections
 \begin{eqnarray}
 \frac{d\sigma}{dt}({Y}) \propto f_Y(\frac{t}{s})
 \left(\frac{1}{s}\right)^{8},
 \label{Eq2}
 \end{eqnarray}
 where $s,t$ are large and the ratio $t/s$ is fixed. Analysis of high-energy
 processes like $\pi p\to \pi p$,  $p p\to p N^*$ etc.,  at fixed values of $t/s$
 shows rather good agreement between the data and power-law
 exponent~\cite{MMT,BF}. This indicates weak energy dependence
 of the function $f(t/s)$ in Eq.~(\ref{Eq2}) which can be considered as
 as a constant. These predictions hold when
 $s$ and $t$ are much larger than the masses of the
 hadrons involved in the processes.

 The dimension of the invariant amplitude $\pi N\to \Theta^+\bar K$
 shown in Fig.~1b is $[\rm length]^{8}$
 which results in
 \begin{eqnarray}
 \frac{d\sigma}{dt}({\Theta^+}) \propto f_{\Theta}(\frac{t}{s})
 \left(\frac{1}{s}\right)^{10}.
 \label{Eq4}
 \end{eqnarray}
 Therefore, for  the ratio $R_{\Theta Y}$, we find
\begin{equation}
 R_{\Theta Y}^{\rm hard} \propto \left(\frac{s_0}{s}\right)^{2}.
\end{equation}
 This estimation has rather illustrative sense because
 the cross sections both for $Y$ and $\Theta^+$
 production decrease steeply with
 $s$. Nevertheless, one can see that the production rate
 of $\Theta^+$ is suppressed at sufficiently high $s$
 as compared to that of the conventional three quark hyperon $Y$.

\section{exclusive production in diffractive region}

\begin{figure}[ht] \centering
    \includegraphics[width=0.8\columnwidth]{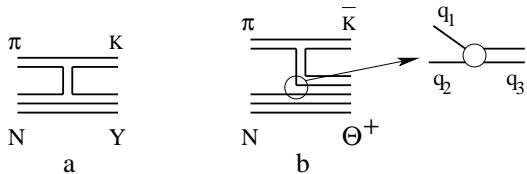}\qquad
 \caption{The hyperon (a) and $\Theta^+$ (b) -production in
 diffractive region.}
 \label{fig:2}
\end{figure}

 The dual quark diagram for the hyperon production
 is shown in Fig.~\ref{fig:2}a.  At low energy this diagram may be
 interpreted as the $t$-channel $K$ and $K^*$ meson exchange processes.
 At large energy they transform to the Regge trajectories where
 the dominant contribution comes from the lowest $K^*$-trajectory
 with the intercept $\alpha_{K^*}\simeq 0.3$~\cite{Regge}.
 This
 results in
 \begin{eqnarray}
  R_Y=\left(\frac{s_0}{s}\right)^{1.4}.
 \end{eqnarray}
 The corresponding dual diagram for the $\Theta^+$ production is shown
 in Fig.~\ref{fig:2}b.  Together with $K$ and $K^*$-exchange
 processes, there is some additional suppression factor
 proportional to the amplitude of picking up the fast moving
 quark-antiquark pair in the projectile (pion) by the slowly moving
 quark(s) in the target (nucleon).
 In the non-relativistic limit, this amplitude is related to
the wave function of the bound quark-antiquark pair
at the large relative momentum ${\bf q}={\bf q}_1 -{\bf q}_2$.
 For the estimation of this effect in the relativistic
 case, one can use the light cone representation for the
 wave functions of the colliding
 hadrons. Neglecting the transverse momentum distributions,
 one can find the following expression
 \begin{eqnarray}
 &&\sqrt{\delta R_{\Theta}}\propto\int\limits_0^1
 dx\int\limits_0^1 dy \,\varphi_\pi(x)\,\varphi_N(y)
 \delta((x+y)^2 - \Delta^2),\nonumber\\
 &&\propto \Delta^{M+N},
 \end{eqnarray}
where $\varphi_{\pi}$ and $\varphi_{N}$ are the light cone  wave
function
 \begin{eqnarray}
  \varphi_\pi(x)\propto x^M(1-x)^M,\,\,\varphi_N(y)\propto
  y^N(1-y)^N,
 \end{eqnarray}
 and
 \begin{eqnarray}
 x\simeq2\frac{p_i}{\sqrt{s}},\qquad
 y\simeq2\frac{p_j}{\sqrt{s}},\qquad
 \Delta\sim\frac{\sqrt{m_q^2}}{\sqrt{s}}.
 \end{eqnarray}
For giving  an upper bound we can choose $M=N=1$ which leads to
\begin{equation}
 R_{\Theta Y}^{\rm diff} \leq \left(\frac{s_0}{s}\right)^{2}.
\end{equation}

\begin{figure}[ht] \centering
    \includegraphics[width=0.3\columnwidth]{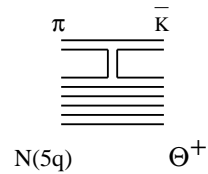}\qquad
 \caption{The $\Theta^+$ production through the 5-quark component
 of the nucleon.}
 \label{fig:3}
\end{figure}

Another estimation can be performed as it is suggested that
$\Theta^+$ may be produced through the five-quark admixture in the
nucleon wave function~\cite{Diakonov04}. The relevant dual diagram
for this process is shown in Fig.~\ref{fig:3}.
Here we have no dynamical suppression factor as discussed above,
but this process is suppressed due to the small probability of the
5-quark component in a nucleon. For a numerical estimation, let us
assume that the isospin and spin-parity of $\Theta^+$ are
$I(J^P)=0(\frac{1}{2}^+)$, and introduce the parameter $\xi$ which
is the amplitude of the 5-quark admixture in a nucleon. Then, for
the $\Theta NK$ and $\Lambda NK$ couplings we can write the ratio
\begin{eqnarray}
|\frac{g_{\Theta NK}}{g_{\Lambda NK}}|= \vert \xi\vert\,
 \vert \frac{\langle\Theta^+|N(5q)K\rangle}{\langle\Lambda|N(3q)\bar
 K\rangle}\vert,
\end{eqnarray}
and therefore,
\begin{eqnarray}
|\xi|\simeq |\frac{g_{\Theta NK}}{g_{\Lambda NK}}|.
\end{eqnarray}
Here we  have taken into account that
 $I(J^P)_\Lambda=0(\frac{1}{2}^+)$ and assumed
 $ {\langle\Theta^+|N(5q)K\rangle}\simeq {\langle\Lambda|N(3q)\bar K\rangle}$.
 For $\Lambda NK$ coupling, one can use the SU(3) relation
 ${g_{\Lambda NK}}=-(3F+D)/\sqrt{3}(F+D){g_{\pi NN}}$
 with $F/D\simeq 0.575$~\cite{F/D}, which gives  ${g_{\Lambda NK}}\simeq-{g_{\pi
 NN}}$. For  $\Theta NK$ coupling, we can use the relation between ${g_{\Theta NK}}$
 and $\Theta$ decay width
 \begin{eqnarray}
 \Gamma_{\Theta^+}=\frac{[g^{}_{\Theta NK}]^2 p_F}{2\pi M_\Theta}
 (\sqrt{M_N^2 + p_F^2}- M_N),
 \end{eqnarray}
where $p_F$ is the $\Theta$ decay momentum, which results in
${g_{\Theta NK}}\simeq 1$ at $\Gamma_\Theta\simeq 1$
MeV~\cite{KN-scattering}. This gives the following estimation
\begin{equation}
 R_{\Theta Y}^{\rm diff}(5q) \simeq 5.6\times 10^{-3}.
\end{equation}

\section{$\Theta^+$-production in fragmentation region}

\begin{figure}[ht] \centering
    \includegraphics[width=0.7\columnwidth]{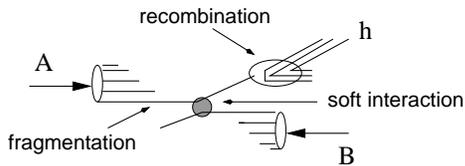}\qquad
 \caption{Production of $\Theta^+$
 in inclusive reactions in the fragmentation region.}
 \label{fig:4}
\end{figure}

 Consider now the $\Theta^+$-production in inclusive reaction
 $AB\to \Theta^+X$ together with the hyperon
 production: $AB\to YX$. The cross section of these reactions
 may be estimated on the base of fragmentation-recombination
 model, which assumes the elementary sub-processes as depicted
 in Fig.~\ref{fig:4}.
Thus, it is assumed that at the first stage the colliding hadrons
fragmentate into partons (quark, gluon, di-quarks, etc). The
probability to find $i$-th constituent (parton) is described by
the "fragmentation" function $F_{i/A}(x)$, where $x=p_i/p_A$. At
the second stage there is some soft (quasi-elastic) interaction
between parton $i$ and some other constituent from the hadron $B$.
Finally, the parton $i$ is recombined into the observable hadron
$h$ ($\Theta^+$ or $Y$). The probability of this process is
defined by the "recombination" function $R_{h/i}(y)$, where
$y=p_h/p_i$. Thus, the cross section of  $AB\to
\Theta^+X$-reaction  is controlled by the folding
\begin{eqnarray}
\sigma\propto\int F_{i/A}(x) R_{h/i}(y) \delta(z-xy)dxdy,
\label{E-1}
\end{eqnarray}
where $z=p_h/p_A$.
By making use of the scale behaviour of $F_{i/A} (x)$
and $R_{h/i}(y)$
\begin{eqnarray}
F_{i/A}(x)\sim F_0(x)(1-x)^{b^i_A};\qquad R_{h/i}(y)\sim
R_{0}(y)(1-y)^{c^h_i},\nonumber
 \end{eqnarray}
where $F_0(x)$ and $R_0(x)$ are smooth functions of $x$, and
keeping the dominant terms we get
\begin{eqnarray}
\sigma\propto\int\limits_z^1 (1-x)^{b^i_A}(x-z)^{c^h_i}d x.
\end{eqnarray}
The integral can be performed by an elementary method, and we can
estimate the cross section as ($b_A^i \equiv b, c_i^h \equiv c$)
\begin{eqnarray}
 \frac{b! c!}{(b+c)!}\frac{1}{b+c+1} (1-z)^{b+c} .
 \label{E-int2}
\end{eqnarray}
For further estimation, we have to specify the power $b$ and $c$
in the fragmentation and recombination functions. In the
quark-parton picture~\cite{Close}, these coefficients are related
to the number of the constituent partons in $A$ and $h$:
$b^i_A=2n_A-3$ and $c_i^h=2n_h-3$. Consider now two extreme
variants. Firstly, we assume the quark-diquark picture of the
hadrons $A$ and $h$. When $A$ is a nucleon and $h$ is a hyperon or
$\Theta^+$ we have $b=1$, $c(\Theta^+)=3$ and $c(Y)=1$. The
corresponding ratio of $\Theta^+$ to $Y$-production reads
\begin{eqnarray}
 {R_{\Theta Y}^{\rm fr.}}_{\rm diquarks} \simeq
\frac{3! 2!}{4!}\frac{3}{5} (1-z)^2 = 0.3(1-z)^2. \label{R1}
\end{eqnarray}
In the quark picture $b=3$, $c(\Theta^+)=7$ and $c(Y)=3$, and
\begin{eqnarray}
 {R_{\Theta Y}^{\rm fr.}}_{\rm quarks} \simeq
\frac{6! 7!}{3! 10!}\frac{7}{11}(1-z)^4\simeq 0.11 (1-z)^4.
\label{R2}
\end{eqnarray}
Combining these extreme cases we get the following estimation
\begin{eqnarray}
 && R_{\Theta Y}^{\rm fr. }\simeq
 \left\{ 0.11(1-z)^4\,\,\,[{\rm quarks}]
 \atop
  ~0.30(1-z)^2\,\,\,[{\rm diquarks}]\right. .
  \label{R3}
\end{eqnarray}
Choosing for $z$ the typical value for the fragmentation region
$z\simeq 0.7$ we get the following bounds
\begin{eqnarray}
 R_{\Theta Y}^{\rm fr. }\simeq
 \left\{ 9 \times 10^{-4} \,\,\,[{\rm
 quarks}]
 \atop
  ~3 \times 10^{-2}\,\,\, [\rm diquarks]\right. .
  \label{R4}
\end{eqnarray}
 which means that the $\Theta^+$-production in the fragmentation region
 is strongly suppressed.
 Notice that  $(1-z)$-power behaviour of the hadron production
 cross sections in the fragmentation region as a rule starts from
 $z\simeq0.4-0.5$~\cite{fragmentregion}.
 At $z=0.5$, the accuracy
 of Eqs.~(\ref{R1}) and (\ref{R2}) is 20 ad 35\%,
 respectively. For $z\simeq0.7$ it is 10 ad 20\%, respectively and
 becomes better when $z\to 1$. At $z\leq 0.4$, we have to specify
 the functions $F_0(x)$ and $R_0(x)$ in Eq.~(\ref{E-1}), which may
 be important for the central rapidity region.
 We also have to include the dependence on the transversal momentum
 (for the finite $p_\perp$) which is, however, beyond the scope of our
 present qualitative analysis.

In summary, we have analyzed the high energy limit of the
$\Theta^+$-pentaquark production. Our consideration is based on
the well-known high energy phenomenology: energy dependence of the
Regge trajectories and the scaling behaviour of the hadronic
amplitudes. We found distinct decreasing of the ratio of the
$\Theta^+$ production compared to the background processes in
diffractive processes  and exclusive reactions with large momentum
transfers. In the fragmentation region at high energy, this ratio
is rather small. Our estimation is done on the base of the
fragmentation-recombination model but it has a general character
and is valid for any model (for example, the relativistic string
model). Physically, the $\Theta^+$-pentaquark production in the
fragmentation region is accompanied by creation of additional 2
quark-antiquark (diquark-antidiquark) pairs with subsequent pick
quarks up by the outgoing hadron. This results in additional
suppression factor $(1-z)^\alpha$ with $\alpha\gtrsim 2$. It may
be worthwhile the point out that there will be no suppression with
increasing energy in the central rapidity regions in inclusive
reactions. Nevertheless, the $\Theta^+$ production at low energies
seems to be most suitable for the study of the properties of
$\Theta^+$.

\acknowledgments

 We thanks H.~Ejiri, K. Hicks, M.~Fujiwara, T.~Nakano,
 and H.~Toki for fruitful discussion. Especially we would like to
 thank D.~Diakonov for discussion of the 5-quark admixture in a nucleon.
 One of authors (A.I.T.) thanks
 T.~Tajima, the director of Advanced Photon Research Center,
 for his hospitality to stay at SPring-8.
 This work is supported in part by the Grant for Scientific
 Research ((C) No.16540252) from the Ministry of Education,
 Culture, Science and Technology, Japan.

\end{document}